\documentclass[]{aastex631}

\shorttitle{Constraints on the Slope of the Dust Attenuation Law}
\shortauthors{Prescott et al.}

\usepackage{color}
\usepackage{amsmath}

\newcommand{\testgal}{GN3 34368}
\newcommand{\paa}{Pa$\alpha$}
\newcommand{\pab}{Pa$\beta$}
\newcommand{\ha}{H$\alpha$}
\newcommand{\hb}{H$\beta$}
\newcommand{\nii}{[\hbox{{\rm N}\kern 0.1em{\sc ii}}]}

\begin{document}

%Full title
\title{Using Multiple Emission Line Ratios to Constrain the Slope of the Dust Attenuation Law}

\correspondingauthor{Moire K. M. Prescott}
\email{mkpresco@nmsu.edu}

\author[0000-0001-8302-0565]{Moire K. M. Prescott}
\affiliation{Department of Astronomy, New Mexico State University, P. O. Box 30001, MSC 4500, Las Cruces, NM, 88003, USA}

\author[0000-0002-0496-1656]{Kristian M. Finlator}
\affiliation{Department of Astronomy, New Mexico State University, P. O. Box 30001, MSC 4500, Las Cruces, NM, 88003, USA}

\author[0000-0001-7151-009X]{Nikko J. Cleri}
\affiliation{Department of Physics and Astronomy, Texas A\&M University, College Station, TX, 77843-4242 USA}
\affiliation{George P.\ and Cynthia Woods Mitchell Institute for Fundamental Physics and Astronomy, Texas A\&M University, College Station, TX, 77843-4242 USA}

\author[0000-0002-1410-0470]{Jonathan R. Trump}
\affiliation{Department of Physics, University of Connecticut, 196A Auditorium Road, Unit 3046, Storrs, CT 06269, USA}

\author[0000-0001-7503-8482]{Casey Papovich}
\affiliation{Department of Physics and Astronomy, Texas A\&M University, College Station, TX, 77843-4242 USA}
\affiliation{George P.\ and Cynthia Woods Mitchell Institute for Fundamental Physics and Astronomy, Texas A\&M University, College Station, TX, 77843-4242 USA}

\begin{abstract}

We explore the possibility and practical limitations of using a three-line approach to measure both the slope and 
normalization of the dust attenuation law in individual galaxies. To do this, we focus on a sample of eleven 
galaxies with existing ground-based Balmer H$\alpha$ and H$\beta$ measurements from slit spectra, plus space-based 
grism constraints on Paschen-$\beta$.  When accounting for observational uncertainties, we show that one galaxy 
has a well-constrained dust law slope and normalization in the 
range expected from theoretical arguments; this galaxy therefore provides an example of what may 
be possible in the future.  However, most of the galaxies are best-fit by unusually steep or shallow slopes. 
We then explore whether additional astrophysical effects or observational biases could 
explain the elevated Paschen-$\beta$/\ha\ ratios driving these results. 
We find that galaxies with high Paschen-$\beta$/\ha\ ratios may be explained 
by slightly sub-unity covering fractions ($>$97\%). 
Alternatively, differing slit losses for different lines can 
have a large impact on the results, emphasizing the importance of measuring all 
three lines with a consistent spectroscopic aperture. 
We conclude that while the three-line approach 
to constraining the shape of the dust attenuation law in individual galaxies 
is promising, deep observations and a consistent observational strategy 
will be required to minimize observational biases and to disentangle 
the astrophysically interesting effect of differing covering fractions. 
The James Webb Space Telescope will provide more sensitive measurements 
of Balmer and Paschen lines for galaxies at $z\approx0.3-2$, enabling 
uniform constraints on the optical-infrared dust attenuation law and its intrinsic variation.
\end{abstract}

%Keywords - use Unified Astronomy Concepts, that will enter during submission process: 
% Astrophysical dust processes(99), Interstellar dust extinction(837), Interstellar dust processes(838), Emission line galaxies(459), Galaxies(573)

\section{Introduction}
\label{sec:intro}

Dust attenuation is an important factor in interpreting observations of galaxies. 
This is particularly true for high-redshift galaxies, for which our primary 
observational constraints are confined to the restframe ultraviolet 
where the effect of dust extinction is particularly strong.  
Our ability to account for dust affects the accuracy of 
derived galaxy properties such as the implied star formation rates, ionization parameters, 
and gas-phase metallicities. Complicating the situation further, studies have uncovered evidence that 
the dust law in galaxies is not universal, neither amongst members of 
the Local Group \citep[e.g.,][]{pei1992,gordon2003} nor at higher redshifts 
\citep[$z\sim0.5-3.0$;][]{kriek2013,salmon2016}. 

Correcting an observed quantity for the effects of dust 
requires understanding both the amount of dust along the line-of-sight 
and the amount of extinction that dust causes as a function of wavelength, 
i.e., the dust extinction law, which is related to the dust composition and 
distribution of grain sizes. 
When considering an entire galaxy, it is often more accurate to 
use the term ``attenuation" to describe the effects of dust \citep[e.g.,][]{calzetti2001,salim2020}.  
Whereas dust extinction is the removal of light from the line-of-sight 
due to absorption and scattering by dust, dust attenuation encompasses the 
contributions of absorption plus scattering both out of and back into of 
the line-of-sight. 

In practice, it is non-trivial to determine the dust attenuation for distant 
galaxies. Typically, the ratio of two nebular lines with a known intrinsic ratio 
is used to estimate the amount of dust attenuation, 
which is then used to scale a dust attenuation law (the curve representing 
dust attenuation versus wavelength) of an assumed shape.
The popular Balmer decrement method, for example, involves the observed flux ratio 
of the \ha$\lambda6563$\AA\ and \hb$\lambda4861$\AA\ lines, while ratios of longer 
wavelength hydrogen lines in the Paschen and Brackett series have also been explored 
for sightlines to nearby star-forming regions \citep{puxley1994,petersen1997,greve2010}.
Commonly used empirical attenuation curves include the prescriptions of \citet{cardelli1989}, 
based on measurements of individual stars in the Milky Way, and \citet{calzetti2000}, 
derived using a sample of local starburst galaxies.  
\citet{charlot2000} showed that a power-law dust attenuation law in which the optical depth $\tau_{\lambda}$ 
is proportional to $\lambda^{-n}$ 
is successful at reproducing observations of starburst galaxies, 
with the exponent $n$ being sensitive to whether the radiation is 
emitted from the diffuse interstellar medium ($n\sim0.7$) versus from stars embedded in their 
birth clouds ($n\sim1.3$). 
This flexible attenuation law has been implemented in spectral energy distribution 
fitting codes \citep[e.g., the Code Investigating GALaxy Emission, CIGALE;][]{boquien2019}

Under physically plausible assumptions about nebular conditions (electron density, $n_{e}$, and temperature, $T_{e}$), 
combining constraints on \ha\ and \hb\ with an additional nebular hydrogen line at a wide wavelength 
separation from the other two, e.g., a Paschen series line in the near-infrared, 
can provide separate measurements of the dust attenuation at two different wavelengths 
and in principle, therefore, constrain both the normalization and 
the slope of the dust attenuation law within an individual galaxy. 
While conceptually straight-forward, the approach has not seen widespread use. 
This is likely due to the fact that Paschen series lines are in the rest-frame 
near-infrared and are intrinsically weaker than the strong Balmer lines in the rest-frame optical. 
As a result, Paschen emission line flux measurements have been published for 
only a handful of galaxies.
For example, HST/NICMOS was used to obtain Paschen-$\alpha$ (\paa; $\lambda=1.875$ $\mu$m) 
imaging of local galaxies \citep{alonsoherrero2006,calzetti2007,kennicutt2007}, 
although these data covered only the central 1\arcmin\ of each galaxy. 
The full disks of two nearby galaxies were imaged more recently in 
Paschen-$\beta$ (\pab; $\lambda=1.2822$ $\mu$m) using HST/WFC3 \citep{kessler2020}. 
Further afield, a recent study combined measurements of ground-based \ha\ and 
\hb\ line measurements with \pab\ constraints obtained with the HST/WFC3 grism 
for eleven galaxies at $z\approx0.2$ \citep{cleri2020}. Some of these galaxies 
showed extreme ratios of \pab\ to \ha, raising a range of possibilities such as 
varying covering fractions, underestimated slit loss corrections, 
or unusual dust laws.
At higher redshifts ($z>2$), \paa\ measurements have been reported for a handful of 
individual lensed galaxies \citep{papovich2009, finkelstein2011, shipley2016}.

In this paper, we explore what constraints on three nebular lines (specifically 
\ha, \hb, and \pab) can tell us about the range of dust laws in the sample 
of eleven $z\approx0.2$ galaxies \citep{cleri2020}.  
In Section~\ref{sec:sample}, we introduce the observed galaxy sample, and 
the emission line fluxes and galaxy morphology measurements 
that will be used in our analysis. 
We explore the implied constraints on dust law slopes and normalizations 
for the galaxy sample using an analytic approach in Section~\ref{sec:analytic}. 
In Section~\ref{sec:grid}, we estimate the implied dust 
law parameters after folding in observational uncertainties, 
and then in Section~\ref{sec:influences}, we explore whether additional 
astrophysical or observational effects could be biasing the observed line ratios. 
We discuss the implications of these results for the utility and reliability 
of the three-line method in Section~\ref{sec:discussion}, and we 
conclude in Section~\ref{sec:conclusions}.  For consistency with 
\citet{cleri2020}, we assume a WMAP9 cosmology 
($\Omega_{M}$=0.287, $\Omega_{\Lambda}$=0.713, $h$=0.693); 
the angular scale at $z\approx0.2$ is 3.3~kpc/\arcsec.  
We also assume the intrinsic emission line ratios appropriate for Case~B 
recombination, i.e., H$\alpha$/H$\beta$=2.86 and \pab/\ha=17.6 ($T=10^4$ K, $n_{e}=10^4$ cm$^{-3}$), 
values that are relatively insensitive to density and gas temperature \citep{ost89}. 
Unless otherwise specified, throughout this paper we use $\tau_{V}$ to refer 
to the nebular attenuation, which is typically a factor of $\sim2$ higher than 
the attenuation of the stellar continuum \citep[e.g.,][]{calzetti1997}.

%%%%%%%%%%%%%%%%%%%%%%%%%%%%%%%%%

\section{The Target Sample}
\label{sec:sample}

\citet{cleri2020} presented a sample of eleven galaxies with flux measurements for 
all three lines -- \pab, \ha, and \hb. The disk-integrated \pab\ flux measurements 
were derived as part of the CANDELS Lya Emission at Reionization (CLEAR) survey \citep[HST GO Cycle 23, PI: Papovich;][]{simons2020}, 
a HST/WFC3 G102 slitless grism survey of 12 pointings in the GOODS-N and 
GOODS-S extragalactic fields \citep{giav04}, with existing multiband HST imaging from the Cosmic Assembly Near-Infrared 
Deep Extragalactic Legacy Survey \citep[CANDELS;][]{grogin11,koekemoer11} and HST/WFC3 G141 grism coverage thanks to 
the 3D-HST Survey \citep{momcheva15,brammer2012}. 
The Balmer line measurements were obtained via ground-based spectroscopy 
from Keck/DEIMOS using a 1\arcsec-wide slit as part of the Team Keck Redshift Survey of the GOODS-N field \citep{wirth2004}. 
DEIMOS does not correct for atmospheric dispersion; we note, however, that the typical 
spatial offset between the \hb\ and \ha\ wavelengths is estimated to be 
$\sim$0.5\arcsec\ (less than the slitwidth) for 
$sec(z)<2$.\footnote{https://www2.keck.hawaii.edu/inst/newsletters/Vol4/Volume\_4.html}
Since these ground-based data were not flux-calibrated, line fluxes were derived by estimating 
the equivalent widths of the lines, interpolating broadband photometric magnitudes of the entire 
galaxy to compute the absolute magnitude in the continuum, and computing the corresponding total galaxy line fluxes \citep{weiner2007}. 
This approach was used to account on average for both throughput and slit losses, under the assumption that the equivalent 
width is constant across the galaxy. No corrections were made for Balmer stellar absorption, since 
at the spectral resolution of DEIMOS, the narrower Balmer emission lines can be distinguished from the 
broader underlying absorption profile. 

For our analysis, we collected the galaxy redshifts and \pab, \ha, and \hb\ line 
fluxes from \citet{cleri2020}, 
as well as the morphological measurements (effective radius $r_{eff}$, 
S\'ersic $n$, axis ratio, position angle) for each galaxy from \citet{vanderwel2012}.

\section{Analytic Approach}
\label{sec:analytic}

With two line ratio measurements (\ha/\hb\ and \pab/\ha) and two parameters (slope and normalization), 
it is possible to constrain a galaxy's nebular attenuation law 
analytically.  We start by assuming a uniform dust screen with no slit losses as our baseline scenario. 
Similar to what is done elsewhere \citep{charlot2000}, we parameterize the 
dust attenuation law using the optical depth as a function of wavelength $\tau_{\lambda}$, scaled 
to the V-band (5500\AA) optical depth $\tau_{V}$:

\begin{equation}
\tau_{\lambda} = \tau_{V} (\lambda/5500\mathrm{\AA})^{-n}
\end{equation}
where $n$ is the dust law slope, and the normalization, $\tau_{V}$, can be expressed as 
$A_{V} = 1.086 \tau_{V}$ in magnitudes.

The observed fluxes of the emission lines can then be written as:
\begin{equation}
\label{eq:pab}
F_{i} = L_{i} e^{-\tau_{V}(\lambda_{i}/5500\mathrm{\AA})^{-n}} \times \frac{1}{4 \pi d_{L}^2}
\end{equation}
where $i=$[\pab, \ha, \hb], and  $d_{L}$ is the luminosity distance at the redshift of the galaxy.

The relevant line ratios are therefore:
\begin{equation}
\label{eq:pabharatio}
\frac{F_{Pa\beta}/F_{H\alpha}}{L_{Pa\beta}/L_{H\alpha}} = e^{-\tau_{V}[(\lambda_{Pa\beta}/5500\mathrm{\AA})^{-n} - (\lambda_{H\alpha}/5500\mathrm{\AA})^{-n}]}
\end{equation}
\begin{equation}
\label{eq:hahbratio}
\frac{F_{H\alpha}/F_{H\beta}}{L_{H\alpha}/L_{H\beta}} = e^{-\tau_{V}[(\lambda_{H\alpha}/5500\mathrm{\AA})^{-n} - (\lambda_{H\beta}/5500\mathrm{\AA})^{-n}]}
\end{equation}

Combining these two equations and eliminating $\tau_{V}$, we derive a relationship between the observable line ratios, the 
known intrinsic line ratios, and the dust law slope $n$:

\begin{equation}
\label{eq:finalratio}
\frac{\ln[(F_{H\alpha}/F_{H\beta})/(L_{H\alpha}/L_{H\beta})]}{\ln[(F_{Pa\beta}/F_{H\alpha})/(L_{Pa\beta}/L_{H\alpha})]} 
=\frac{ (\lambda_{H\beta}/\lambda_{H\alpha})^{-n} - 1}{1 - (\lambda_{Pa\beta}/\lambda_{H\alpha})^{-n}}
\end{equation}

Solving Equation~\ref{eq:finalratio} using Newton-Raphson 
iteration, we obtain $n$; plugging the result back into 
Equations~\ref{eq:pabharatio}-\ref{eq:hahbratio}, we can solve for $\tau_{V}$. 

\begin{figure}
\center
\includegraphics[angle=0,width=7in]{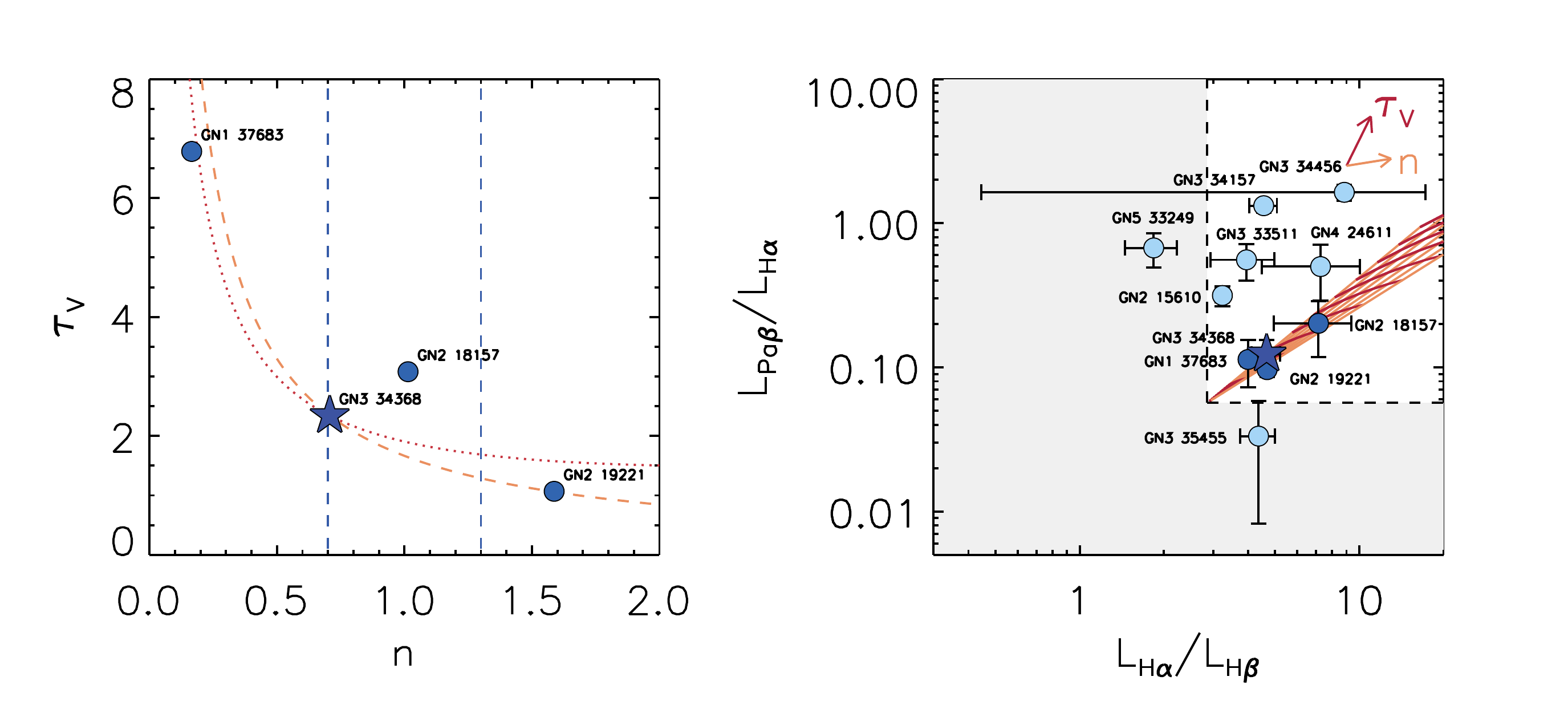}
\caption{(Left) Dust law slopes and normalizations implied by the analytic approach 
for galaxies with acceptable values (non-negative $n$ and $\tau_{V}$ with SFR$<$10 M$_{\odot}$ yr$^{-1}$). 
For one galaxy (\testgal; star), we show the family of solutions from the individual Pa$\beta$/H$\alpha$ 
(red dotted line) and H$\alpha$/H$\beta$ (orange dashed line) 
ratios separately, with the joint solution where the two dotted lines cross. 
Standard dust law slopes expected for diffuse ISM ($n=0.7$) and stellar birth 
clouds ($n=1.3$) are shown \citep[light blue dashed lines;][]{charlot2000}. 
(Right) Observed emission line ratios for the eleven galaxies in the \citet{cleri2020} 
sample (blue filled symbols). 
The expected line ratios for standard dust law slopes (orange; $n=0.7-1.3$) and a range of normalizations 
(red; $\tau_{V}=0.1-10$) are shown, with the arrows indicating the direction of increasing values. 
The intrinsic line ratios, assuming Case~B recombination, are shown with dashed black lines; 
two galaxies are $\sim$1-3$\sigma$ outliers, with measurements lying below these values, 
i.e., in the ``unphysical" regime (grey shaded region). 
The four galaxies that are well-explained by a simple power-law dust 
attenuation law are shown in both panels (filled dark blue symbols), 
with the example galaxy indicated (\testgal; star). Individual galaxy ID numbers are shown in both panels. 
}
\label{fig:fiduciallineratios}
\end{figure}

As a demonstration, we apply this approach first to one of the galaxies 
in the CLEAR sample (\testgal).  The measurement of the Balmer decrement yields 
a curve within the slope-normalization parameter space. The \pab/\ha\ ratio 
corresponds to a second curve, and the point where the two curves cross represents 
the value of slope and normalization implied by the nebular 
lines from that galaxy.  This is shown in Figure~\ref{fig:fiduciallineratios} (left panel).  

When we apply this approach to the rest of the CLEAR sample, we find that 
a total of four galaxies yield physically plausible, albeit varied, dust law slope and 
normalization values.  In Figure~\ref{fig:fiduciallineratios} (left panel), we show the 
results for these four galaxies. For comparison, we also plot the dust law slopes 
proposed by \citet{charlot2000}, with $n=0.7$ representing diffuse interstellar medium 
(ISM) and $n=1.3$ representing stellar birth clouds.  
Hereafter, we refer to the range $n=0.7-3$ as ``standard'' dust laws.  
Under our baseline scenario, the results in Figure~\ref{fig:fiduciallineratios} (left panel) seem to imply 
a wide range of dust laws for this small subset of galaxies. 

In the rest of the galaxy sample, however, this analytic approach 
yields negative dust law slopes and/or normalizations. 
In Figure~\ref{fig:fiduciallineratios} (right panel), we plot 
the observed emission line ratios for the entire sample, along with predictions 
for standard dust law slopes and a range of normalizations. 
Four galaxies have line ratios that are consistent with standard dust laws to within $1\sigma$, 
but a comparable number (five galaxies) have significant \pab/\ha\ excesses. 
While this could in part be due to the fact that this sample was \pab-selected, 
we note that four of these cases 
are more than 3$\sigma$ away from the expectations for standard dust laws. 
Finally, two galaxies are 1-3$\sigma$ ``outliers,''  with observed line ratios 
below the intrinsic Case~B values. 

\section{Accounting for Observational Uncertainties}
\label{sec:grid}

\begin{figure}
\center
\includegraphics[angle=0,width=5.5in]{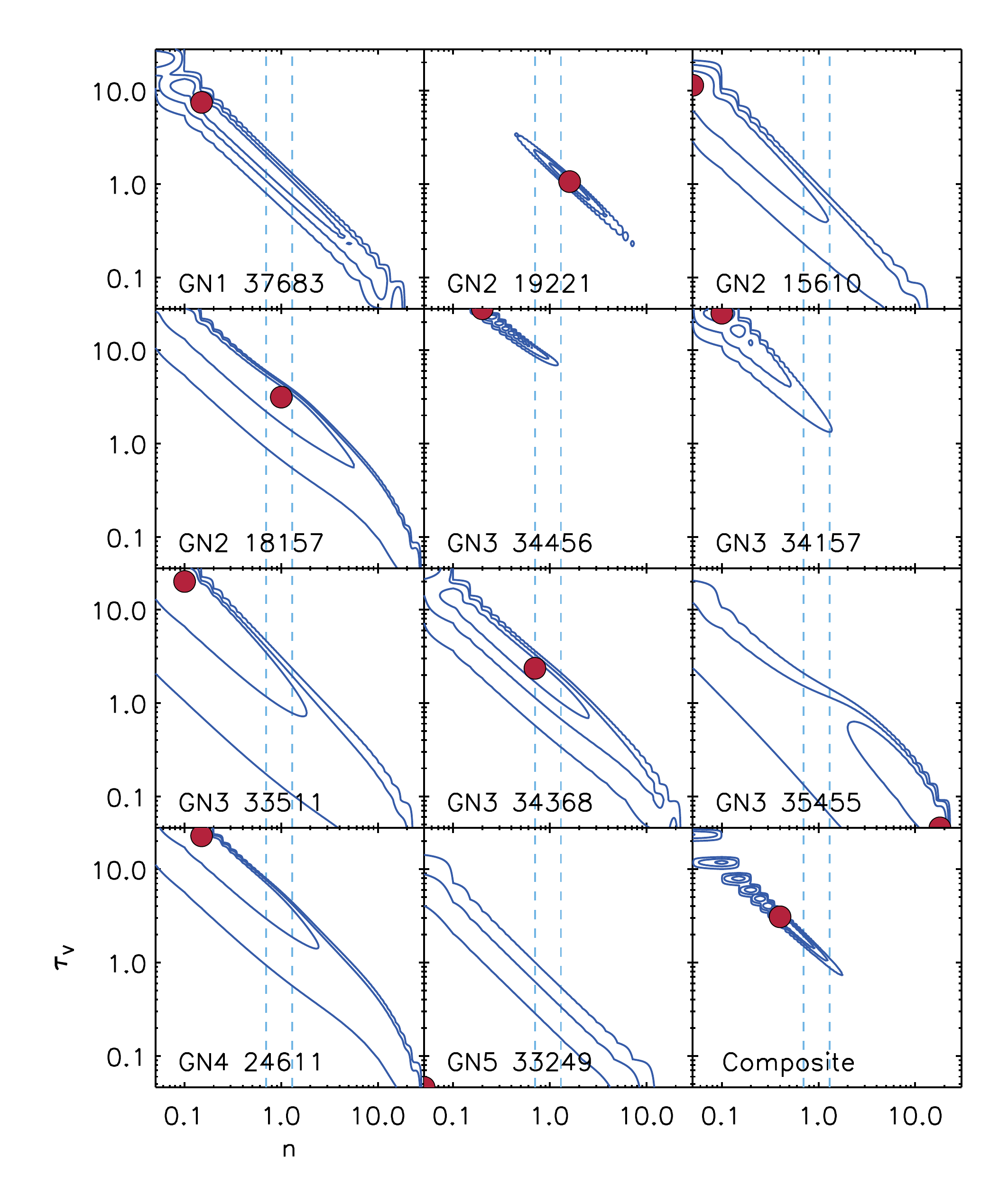} 
\caption{Best-fit dust law slope ($n$) and normalization ($\tau_{V}$) accounting 
for observational uncertainties for each galaxy, and in the lower right panel, for a composite generated by computing the 
error-weighted mean and uncertainty of the full sample. 
Contours contain 68.3\%, 95.4\%, and 99.7\% of the total likelihood given the assumed priors, and the corresponding location 
of the minimum $\chi^2$ is shown (filled red circles). Standard dust law slopes expected for diffuse ISM ($n=0.7$) 
and stellar birth clouds ($n=1.3$) are shown \citep[light blue dashed lines;][]{charlot2000}. 
One galaxy (GN2 19221) is well-constrained with the existing data. Most of the sample overlaps standard 
dust law slopes, given the current observational uncertainties, although some individual galaxies 
(and the full-sample composite) are more consistent with steeper or shallower dust laws.
}
\label{fig:gridplot}
\end{figure}

Assuming our baseline scenario, we can then determine the best estimate for the dust law slope and normalization 
after accounting for observational uncertainties. To do this, we employed a 
posterior-sampling approach, equivalent to a Bayesian inference method with uniform priors. 
We generated a model grid varying the slope and normalization of 
the dust attenuation law.  Assuming the intrinsic ratios 
of \ha/\hb\ and \pab/\ha\ appropriate for 
Case~B recombination, we then calculated the expected observed 
ratios for each combination of dust law slope and normalization.  
We assumed wide uniform priors on both the slope and normalization ($0<n<30$, $0<\tau_{V}<30$). 
For each galaxy, we then used the observed line ratios and observational errors to 
compute $\chi^{2}$, and found the maximum likelihood solution implied by the grid. 
The likelihood contours are shown for each galaxy in Figure~\ref{fig:gridplot}, and the 
implied slope and normalization corresponding to the maximum likelihood 
(minimum $\chi^{2}$) is indicated. 

Out of the sample, one galaxy (GN2 19221) has a well-constrained dust law slope ($n\sim1.3$) and 
normalization ($\tau_{V}\sim1.0$). 
Three additional galaxies (GN1 37683, GN2 18157, GN3 34368) have solutions consistent 
with standard dust law slopes 
(within the contour containing 68.3\% of the total likelihood, given the assumed priors) 
and far from the assumed prior boundaries. These are the same four galaxies 
that yielded acceptable solutions using the simple analytic approach 
(Section~\ref{sec:analytic}). 
However, six galaxies show results at or near the boundary of our uniform prior ranges. 
In three of these cases (GN2 15610, GN3 33511, GN4 24611), 
the 68.3\% confidence intervals overlap standard dust law slopes, but 
for three galaxies (GN3 34456, GN3 34157, GN3 35455), 
one of which (GN3 34456) has the most extreme \pab/\ha\ ratio of the sample, 
the confidence intervals are skewed towards shallower or steeper slopes.  
As before, for one galaxy (GN5 33249), 
the smallest in the sample, no acceptable solution is found. 
If we were to make the assumption that this sample of galaxies 
represents eleven independent realizations of the same underlying physical conditions, we can 
repeat the analysis using a composite galaxy defined by the error-weighted mean and 
uncertainty of the emission line ratios for the full sample. The result, shown in 
the lower right panel of Figure~\ref{fig:gridplot}, implies that this galaxy sample 
prefers somewhat shallower-than-standard dust law slopes.
This is consistent with the take-away from Figure~\ref{fig:fiduciallineratios} (right panel), where 
a fair fraction of the sample showed enhanced \pab/\ha\ ratios above the standard dust law curves, 
i.e., in the direction of lower $n$.
Thus, we are left with the impression that the observed emission line ratios imply a diversity of rather extreme and in some cases 
unphysical dust law slopes across this sample of galaxies.

\section{Influence of Astrophysical versus Observational Effects}
\label{sec:influences}
It is possible that the baseline scenario is not accurate and that the unusual observed line ratios 
driving these results reflect sub-unity covering fraction of the dusty ISM or ground-based Balmer line measurements 
that are either over- or undercorrected for slit losses. 
In the rest of this section, we explore these two possibilities to see how well 
we can explain the line ratios from the observed sample. 

\subsection{Covering Fractions}
\label{sec:coveringfractions}

One explanation for the unusual observed line ratios in certain galaxies is 
different covering fractions of dusty ISM \citep{cleri2020}, such that some regions 
are completely opaque to the Balmer lines whereas others are much less so. 
In this scenario, the disk-integrated Paschen emission comes from the entire galaxy, 
but the Balmer lines are detected predominantly from the subset of sightlines with low dust. 

To explore this effect, we make the following adjustments to the analytic calculation.  
We assume that lines-of-sight within the galaxy are either dusty or dust-free.  
All Paschen and Balmer line flux from the dust-free lines-of-sight is observed, 
whereas line flux from the dusty portion is attenuated according to the dust law. 
We define the covering fraction $f_\mathrm{cov}$ to be the fraction of the galaxy covered by 
dusty ISM. Thus, $f_\mathrm{cov}=0.0$ implies that none of the galaxy experiences dust attenuation, 
whereas $f_\mathrm{cov}=1.0$ means that the entire galaxy experiences dust attenuation given by 
$\tau_{\lambda}$, at which point the situation reduces to the baseline scenario in Section~\ref{sec:analytic}.
We write down the observed line flux ($F_{\lambda}$) as a 
function of the true line luminosity ($L_{\lambda}$):

\begin{equation}
\label{eq:fcovequation}
F_{\lambda} = [(1-f_\mathrm{cov}) L_{\lambda} + f_\mathrm{cov} L_{\lambda} e^{-\tau_{\lambda}}] \times \frac{1}{4 \pi d_{L}^2} 
\end{equation} 

where $\lambda$ corresponds to the wavelengths of \pab, \ha, and \hb, respectively. 
The corresponding line ratios are therefore:

\begin{equation}
\label{eq:fcovratioPabHa}
\frac{F_{Pa\beta}/F_{H\alpha}}{L_{Pa\beta}/L_{H\alpha}} = \frac{(1-f_\mathrm{cov}) + f_\mathrm{cov} e^{-\tau_{V}(\lambda_{Pa\beta}/5500\mathrm{\AA})^{-n}}}{(1-f_\mathrm{cov}) + f_\mathrm{cov} e^{-\tau_{V} (\lambda_{H\alpha}/5500\mathrm{\AA})^{-n}}}
\end{equation}
\begin{equation}
\label{eq:fcovratioHaHb}
\frac{F_{H\alpha}/F_{H\beta}}{L_{H\alpha}/L_{H\beta}} = \frac{(1-f_\mathrm{cov}) + f_\mathrm{cov} e^{-\tau_{V}(\lambda_{H\alpha}/5500\mathrm{\AA})^{-n}}}{(1-f_\mathrm{cov}) + f_\mathrm{cov} e^{-\tau_{V} (\lambda_{H\beta}/5500\mathrm{\AA})^{-n}}}
\end{equation}

Thus far, we have not imposed any restrictions on the analysis, other than 
that the dust law slope and normalization must be non-negative. In practice, however, 
large attenuation of the Balmer lines would imply very 
large, and potentially implausible, star formation rates. 
The measured star formation rates from \citet{cleri2020} 
are all $\lesssim$2 M$_{\odot}$ yr$^{-1}$ based on the observed ultraviolet and infrared luminosities. 
Therefore, we impose a conservative upper limit on the implied star formation rate of 10 M$_{\odot}$ yr$^{-1}$. 
We solve Equations~\ref{eq:fcovratioPabHa}-\ref{eq:fcovratioHaHb} 
and determine the range of covering fractions for which such ``acceptable'' solutions exist 
{(i.e., a non-negative dust law normalization and slope, and an implied dust-corrected star formation rate 
of $<$10 M$_{\odot}$ yr$^{-1}$)}. 

\begin{figure}
\center
\includegraphics[angle=0,width=5.5in]{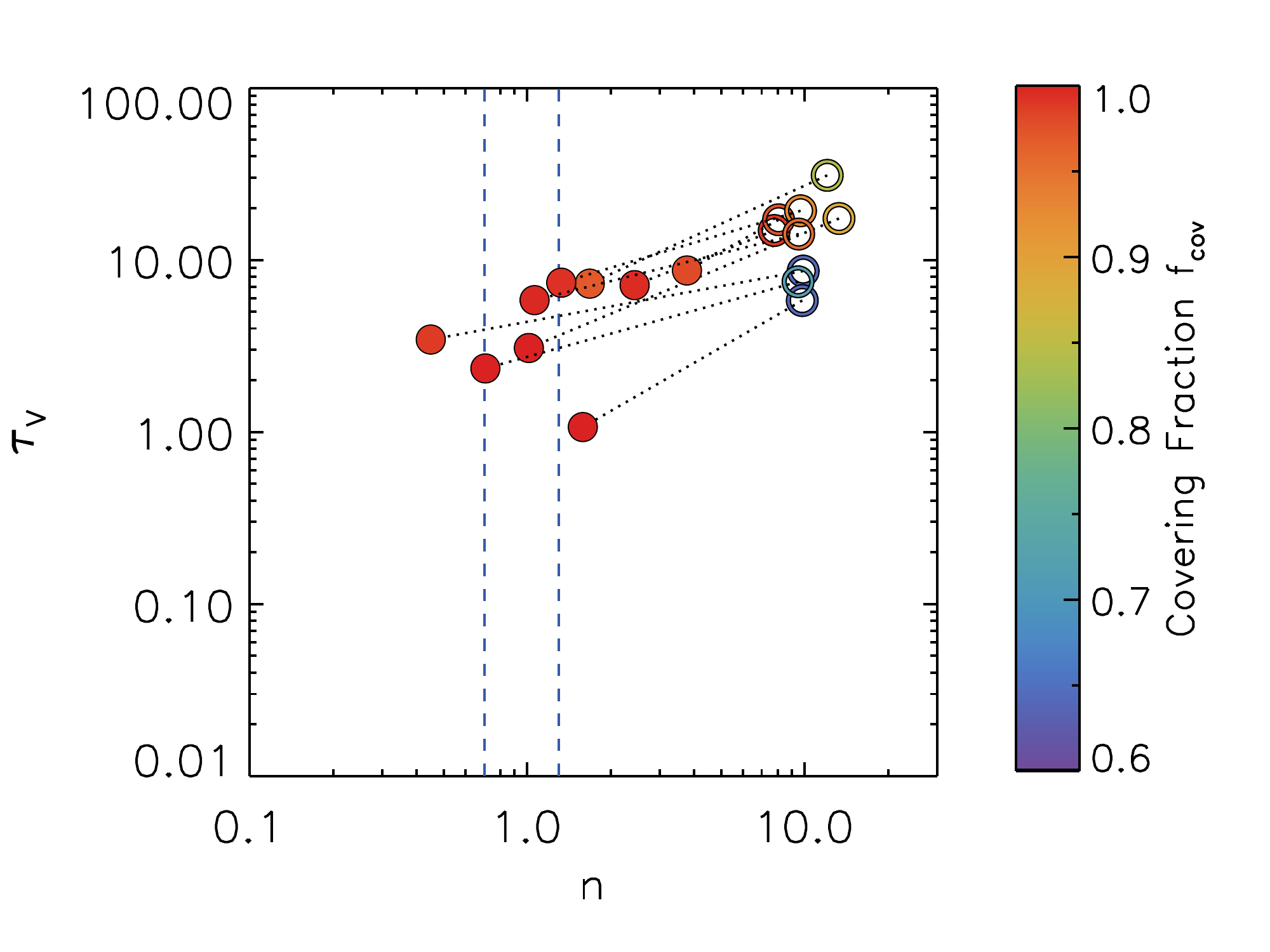}
\caption{Range of covering fractions ($f_\mathrm{cov}$, indicated with the color bar) that produce acceptable results from 
the analytic approach (Equations~\ref{eq:fcovratioPabHa}-\ref{eq:fcovratioHaHb}). Dotted lines connect the minimum (open circle) and maximum (filled circle) 
covering fraction allowed for each galaxy. Standard dust law slopes expected for diffuse ISM ($n=0.7$) 
and stellar birth clouds ($n=1.3$) are shown \citep[light blue dashed lines;][]{charlot2000}.  
While allowing for $f_\mathrm{cov}<1$ permits an acceptable solution for 9/11 galaxies in the sample, 
substantially sub-unity covering fractions imply very extreme dust law slopes ($n>1.3$). 
For standard dust law slopes, high but in some cases sub-unity covering fractions 
($0.97<f_{cov}<1$) are indicated.  
}
\label{fig:coveringfraction}
\end{figure}

Table~\ref{tab:coveringfraction} lists the derived minimum and maximum covering 
fractions for the galaxy sample. 
For two galaxies, there is no acceptable solution.
This is not surprising, as these two galaxies 
are the known outliers, with observed line ratios in the ``unphysical'' regime 
(Figure~\ref{fig:fiduciallineratios}, right panel). 
In Figure~\ref{fig:coveringfraction}, we show the range of covering fractions 
that produce acceptable results for each of the other nine galaxies.  
Allowing for a sub-unity covering fraction can explain the line ratios for the rest of the sample, 
with the lowest allowed covering fractions being 64-99\% and the highest allowed covering fractions being 97-100\%. 
However, in each galaxy, a lower covering fraction implies a steeper dust law slope and higher normalization. 
Since substantially sub-unity covering fractions 
would imply very extreme dust laws, 
they are therefore not the preferred explanation for the line ratios in this sample.  
For standard dust law slopes, 
high but in some cases sub-unity covering fractions ($\gtrsim$97\%) are required by our analysis. 

\subsection{Slit Loss Correction Underestimates}
\label{sec:slitlosses}

On the other hand, observational biases may be driving the unusual line ratios within 
the sample. While the \pab\ measurements represent disk-integrated values, the \ha\ and \hb\ fluxes 
were derived using a spectroscopic slit and were then corrected for both throughput and slit losses 
as described in Section~\ref{sec:sample}. However, this correction assumed that there was no difference in 
equivalent width between regions inside and outside of the slit. 
While this assumption was reasonable on average given a typical galaxy size of $r_{eff} < 0.95$\arcsec\ \citep{weiner2007}, 
it may be violated in galaxies that are large compared to the size of the slit or 
with line emission that is offset or more extended than the continuum. The resulting variable equivalent width 
could lead to larger slit losses for line emission than for continuum.  In that case, the assumption of constant 
equivalent width across the galaxy would result in an underestimate of the emission line slit loss correction 
for the Balmer line fluxes and, as a result, a measured \pab/\ha\ enhancement. 
Figure~1 (right panel) shows that at least half of the galaxy sample have a \pab/\ha\ enhancement 
that cannot be explained by acceptable dust laws, 
therefore it seems plausible that the assumption of constant equivalent width versus position may 
not be valid for some of these galaxies.  

To assess the relative importance of slit losses, we empirically estimated the effective slit loss correction that 
was applied previously (as described in Section~\ref{sec:sample}) using the known size and morphology of the galaxy \citep{vanderwel2012} and the width of the spectroscopic 
slit employed in the ground-based observations \citep[1\arcsec;][]{wirth2004}.  
One complication is that while the ground-based observations used a Keck/DEIMOS multislit mask 
at a particular position angle, the individual slitlets were milled at an orientation within 
$\pm$30 degrees of that angle. Although the stated goal was to align the slit with the major axis 
of the galaxy, when possible, the resulting position angle of the Keck/DEIMOS slitlets 
relative to each galaxy's major axis is not specified.  We therefore computed the slit losses 
for both the best (and more likely) case, with the slitlet fully aligned with the galaxy 
major axis resulting in minimal slit losses, and the worst case, with the slitlet 
perpendicular to the galaxy major axis resulting in maximal slit losses.  
In all cases, we assumed the galaxy was centered in the slitlet.  
While these empirical slit loss corrections could be underestimated, for example, in galaxies 
that depart from a single Sersic profile due to strong dust lanes, they nonetheless provide 
an indication of which galaxies will be most susceptible to errors in the slit loss correction.

The empirically estimated slit losses, along with the tabulated $r_{eff}$ values \citep{vanderwel2012}, 
are given in Table~\ref{tab:empiricalslitlosses}.  
In the best case, the slit losses were between 0-49\%, while in the worst case they ranged from 2-68\%. 
Some galaxies are small compared to the slit (at least as measured by continuum $r_{eff}$), resulting 
in minimal empirical slit loss corrections. 
Interestingly, both of the two outlier galaxies had essentially 
no slit losses due to their small effective radii ($r_{eff}\sim0.2$\arcsec). 
On the other hand, five of the galaxies are large compared to the slit, 
resulting in large ($>$50\%, in the worst case) empirical slit loss corrections. 
This suggests that variable equivalent width or line emission that is more extended than or 
offset from the continuum could plausibly have led to an underestimate in the slit loss correction in these cases.

To further explore the possibility of over- or underestimated slit loss corrections in this sample, we make the following 
adjustments to the analytic calculation.  We use 
$f_\mathrm{slit}$ as a factor that accounts for any additional uncorrected slit losses, with a range from 
-0.99 to 0.99.  Physically, a negative value of 
$f_\mathrm{slit}$ corresponds to the situation where the emission line equivalent width is smaller outside slit than inside, 
resulting in a previous overcorrection, whereas a positive $f_\mathrm{slit}$ means the 
line equivalent width is larger outside the slit, leading the previous correction to be underestimated. 
We then write down the tabulated line flux ($F_{\lambda}$, including the nominal slit loss correction) as a function of 
the true line luminosity ($L_{\lambda}$, fully slit-loss-corrected) for the two Balmer lines:

\begin{equation}
\label{eq:haslit}
F_{H\alpha} = L_{H\alpha} e^{-\tau_{V}(\lambda_{H\alpha}/5500\mathrm{\AA})^{-n}}  \times \frac{(1-f_\mathrm{slit})}{4 \pi d_{L}^2} 
\end{equation} 

\begin{equation}
\label{eq:hbslit}
F_{H\beta} = L_{H\beta} e^{-\tau_{V}(\lambda_{H\beta}/5500\mathrm{\AA})^{-n}}  \times \frac{(1-f_\mathrm{slit})}{4 \pi d_{L}^2} 
\end{equation}

We continue to use Equation~\ref{eq:pab} for \pab, as the disk-integrated slitless grism measurements are unaffected by slit losses. 

The corresponding \pab/\ha\ line ratio is therefore:

\begin{equation}
\label{eq:pabharatioslit}
\frac{F_{Pa\beta}/F_{H\alpha}}{L_{Pa\beta}/L_{H\alpha}} = \frac{e^{-\tau_{V}[(\lambda_{Pa\beta}/5500\mathrm{\AA})^{-n} + (\lambda_{H\alpha}/5500\mathrm{\AA})^{-n}]}}{(1-f_\mathrm{slit})}
\end{equation} 

The \ha/\hb\ ratio is still given by Equation~\ref{eq:hahbratio} as the slit losses are assumed 
to be the same and therefore cancel out for the two Balmer lines. 

\begin{figure}
\center
\includegraphics[angle=0,width=5.5in]{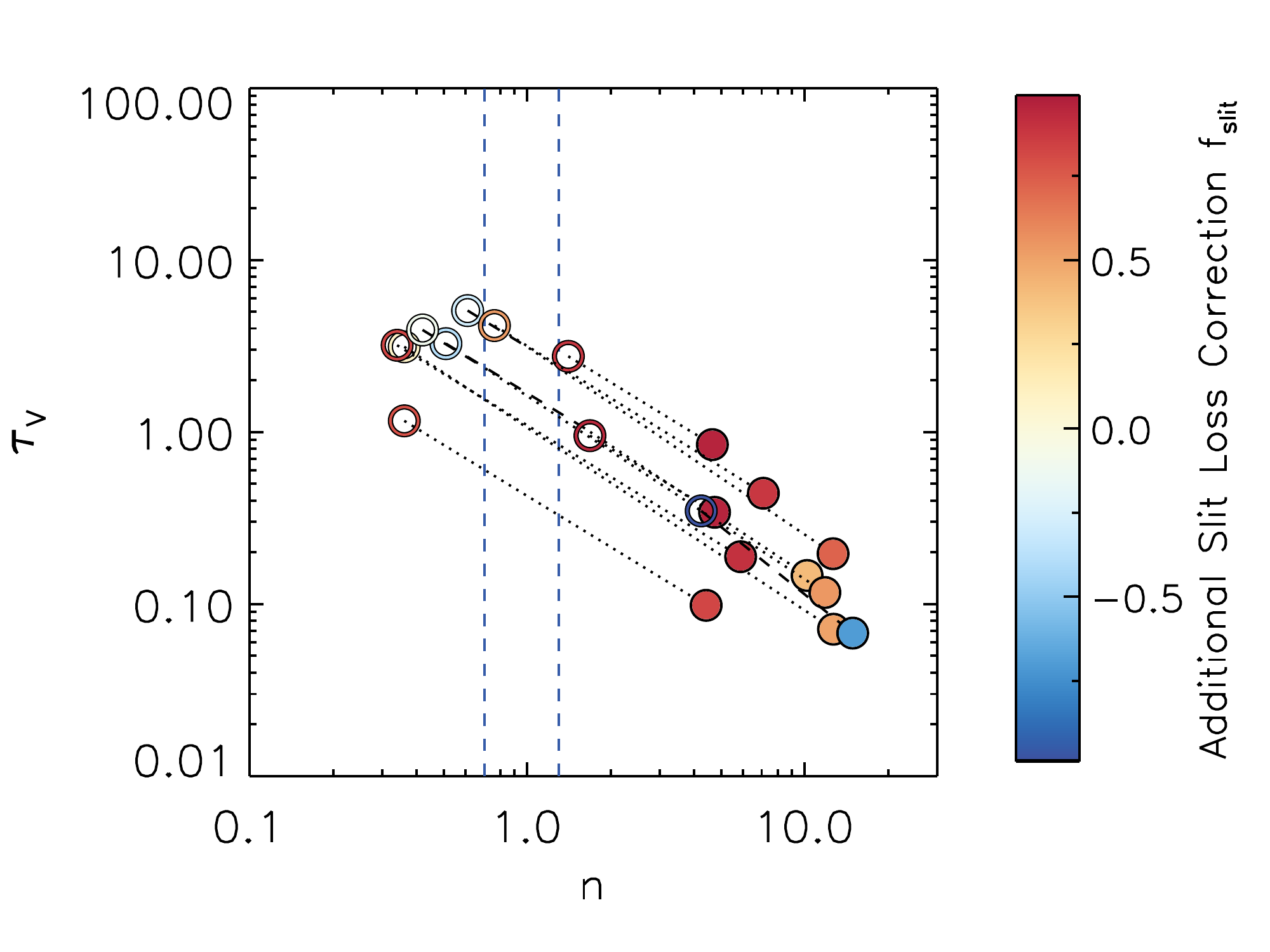}
\caption{(Left) Range of allowed additional slit loss corrections ($f_\mathrm{slit}$, indicated with the color bar) 
that produce acceptable results, i.e., 
non-negative $n$ and $\tau_{V}$ with SFR$<$10 M$_{\odot}$ yr$^{-1}$, 
from the analytic approach (Equations~\ref{eq:hahbratio} and \ref{eq:pabharatioslit}). 
Dashed and dotted lines connect the minimum (open) and maximum (filled) 
$f_\mathrm{slit}$ values for each galaxy, where dashed lines indicate negative $f_\mathrm{slit}$ values (a previous overcorrection) 
and dotted lines indicate positive $f_\mathrm{slit}$ values (a previous undercorrection). 
Standard dust law slopes expected for diffuse ISM ($n=0.7$) and stellar birth 
clouds ($n=1.3$) are shown \citep[light blue dashed lines;][]{charlot2000}. 
Moderate slit loss over- or undercorrection may be present in the tabulated Balmer fluxes, 
assuming that standard dust laws apply.
}
\label{fig:slitlosses}
\end{figure}

We then calculate what range of $f_\mathrm{slit}$ values would be required in order to produce an acceptable 
solution. 
As before, we require non-negative $n$ and $\tau_{V}$, and impose a limit on the implied star formation 
rate of 10 M$_{\odot}$ yr$^{-1}$.  The results are shown in Figure~\ref{fig:slitlosses} 
and in Table~\ref{tab:slitlossunderestimates}, along with the estimated stellar masses from \citet{skelton14}. 
One of the outlier galaxies (GN5 33249) returned unphysical results as before; this is not surprising as the \ha/\hb\ ratio for 
this galaxy is below the Case~B intrinsic value, so there is no slit loss correction that can bring it into the physical regime.
The other outlier galaxy (GN3 35455) 
can be explained but only in the case of a high slit loss correction 
overestimate ($f_\mathrm{slit}<-0.71$), which is implausible given its small physical size ($r_{eff}\sim0.22$\arcsec).
The remaining nine galaxies show a range of additional slit loss corrections.
In three cases, allowable $f_\mathrm{slit}$ values can be either positive or negative, whereas in six cases 
only $f_\mathrm{slit} > 0$ (a slit loss correction underestimate) is permitted. 
Correcting for such slit loss underestimates generically reduces the \pab/\ha\ flux ratio 
towards the intrinsic Case~B value without modifying the Balmer decrement, 
changing the best-fit slope and normalization. 
Thus, correcting for slit losses that may have been moderately over- or underestimated previously 
may bring most, although not all, of the galaxy sample into agreement with standard dust laws.

\section{Discussion}
\label{sec:discussion}

Using an analytic approach, we showed that four of the eleven galaxies are consistent with 
a simple power-law dust attenuation law. However, the required slopes ($n$) ranged 
from much flatter than expected for the diffuse ISM ($n=0.7$) to much steeper than expected 
for stellar birth clouds ($n=1.3$). A closer look revealed that 
the remaining seven galaxies have unusual line ratios that are far from the predicted locus using 
standard dust law slopes. 
While the current sample of galaxies may be biased towards higher \pab/\ha\ and lower \ha/\hb\ ratios 
due to the fact that it was selected by requiring detections in \pab, \ha, and \hb\ \citep{cleri2020}, 
it nonetheless provides a useful opportunity to explore the physical interpretation of these 
more extreme line ratios.  

\subsection{Astrophysical and Observational Considerations}
\citet{cleri2020} discussed the possibility that these unusual line ratios are the result of 
sub-unity covering fractions of dusty ISM.  
This might occur, for example, if a galaxy has a prominent dust lane \citep{cleri2020} 
or a collimated, dusty outflow \citep[e.g.,][]{walter2002}.
In this case, relatively dust-free lines-of-sight would contribute Paschen-to-Balmer line ratios close to their intrinsic Case~B values, 
but dusty lines-of-sight would show highly attenuated Balmer relative to 
Paschen emission, creating an enhancement of the overall observed 
Paschen-to-Balmer line ratio.  

\begin{figure}
\center
\includegraphics[angle=0,width=5.5in]{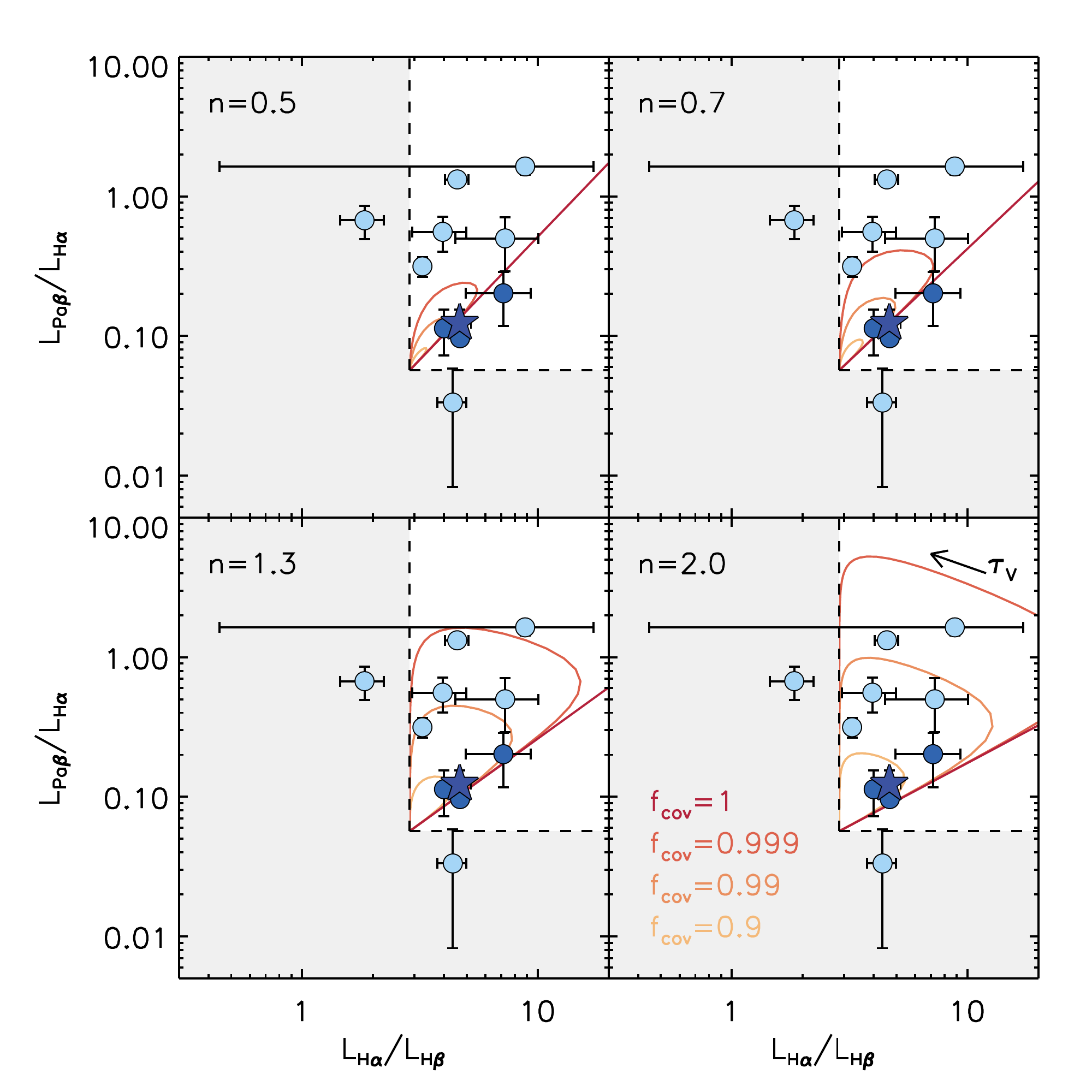}
\caption{Effect of covering fraction ($f_{cov}$) on the \pab/\ha\ versus \ha/\hb\ line ratios 
(orange/red lines), for a different value of the dust attenuation law slope ($n$) in each panel.
The direction of increasing dust attenuation law normalization ($\tau_{V}$) is indicated with an arrow 
in the lower right ($n=2.0$) panel.
For comparison, the observed emission line ratios for the eleven galaxies in the \citet{cleri2020}
sample are shown as in Figure~\ref{fig:fiduciallineratios} (blue filled symbols).
The intrinsic line ratios, assuming Case~B recombination, are shown with dashed black lines, 
with the ``unphysical" regime indicated (grey shaded region). Slightly sub-unity covering 
fractions ($f_{cov}\gtrsim0.9$) lead to extreme line ratios, similar to what is seen in the 
galaxy sample. However, for substantially sub-unity covering fractions ($f_{cov}\lesssim0.9$) 
the emission line ratios return to the Case~B intrinsic values due to the higher fraction 
of unattenuated sightlines.
}
\label{fig:hookfigure}
\end{figure}

\begin{figure}
\center
\includegraphics[angle=0,width=5.5in]{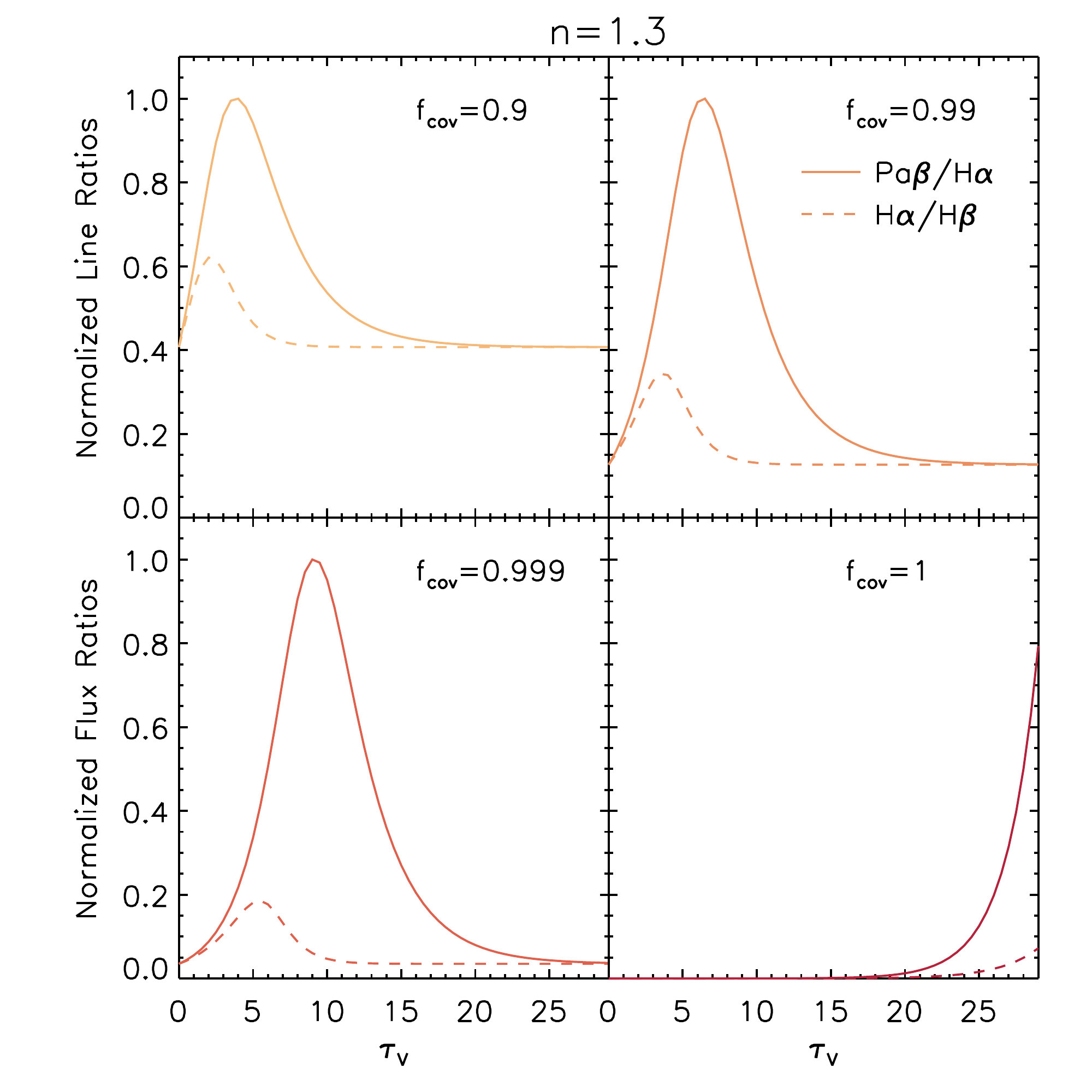}
\caption{Normalized line ratios as a function of dust attenuation law normalization ($\tau_{V}$), 
for different values of covering fraction ($f_{cov}$). 
\pab/\ha\ and \ha/\hb\ ratios are shown as solid and dashed lines, respectively. 
For slightly sub-unity covering fractions ($f_{cov}\gtrsim0.9$), both line ratios 
increase with increasing dust attenuation up to a certain maximum value, after 
which they decline back to their intrinsic values. 
However, the shorter wavelength pair, \ha/\hb, reaches its peak ratio at lower 
dust attenuation than \pab/\ha. Thus, slightly sub-unity covering fractions combined 
with moderate to high dust attenuation can lead to unusually high \pab/\ha\ ratios, as 
seen in the observed sample. 
}
\label{fig:lineratiosVtauv}
\end{figure}

Our analysis has suggested that sub-unity covering fractions could plausibly explain the extreme 
\pab/\ha\ line ratios in at least some of the galaxy sample. 
To further explore how this works, in Figure~\ref{fig:hookfigure}, 
we show the resultant line ratios for four different covering fractions $f_{cov}$, along with 
the observed line ratios for the galaxy sample.  In each panel, we assume a different dust attenuation 
law slope $n$, ranging from shallower ($n=0.5$) to steeper ($n=2.0$) than the standard 
range ($n=0.7-1.3$). The resultant curves correspond to a range of dust law 
normalizations ($0<\tau_{V}<30$), with the direction of increasing $\tau_{V}$ indicated 
with an arrow in the bottom righthand panel.  
With a unity covering fraction, the line ratios increase 
with normalization, as higher dust column leads to progressively greater attenuation of 
the shorter wavelength \ha\ and \hb\ emission lines.  
For slightly sub-unity covering fractions ($f_{cov}\gtrsim0.9$), however, the 
curves form loops in the line 
ratio diagram.  This comes about because at sufficient dust columns, the 
few lines-of-sight that are dust-free start to dominate the observed flux, causing the observed 
line ratios to converge back towards the intrinsic Case~B values.  The dust law 
normalization where this occurs, however, depends on the line ratio, with 
the shorter wavelength \ha/\hb\ ratio beginning to return to intrinsic values at 
lower dust columns than the longer wavelength \pab/\ha\ ratio.  This can be seen in 
Figure~\ref{fig:lineratiosVtauv}, where for non-unity covering fractions, the line 
ratios initially increase as a function of dust law normalization, but then decrease again 
past a critical value, with the \pab/\ha\ ratio turning over 
at higher normalizations than the \ha/\hb\ ratio. 
We note that this effect occurs only for very high but non-unity covering fractions; at 
$f_{cov}\lesssim0.9$, the larger fraction of dust-free lines-of-sight causes the 
observed line ratios to remain close to their intrinsic Case~B values. 
Thus, sub-unity covering fractions do seem to be a plausible explanation for 
most of the unusual line ratios observed in this sample. 
We note that a high or even unity covering fraction is favored for the galaxy with the most extreme 
\pab/\ha\ ratio (GN3 34456), which shows a clear dust lane in the existing GOODS-N 
and CANDELS imaging. Future measurements with more than three emission lines will 
allow for independent constraints on the covering fraction of dusty ISM in galaxies.

A second possibility was that underestimates in ground-based slit loss corrections suppressed the measured 
Balmer line emission, which, when compared with space-based full-galaxy \pab\ measurements, 
resulted in an enhancement of the observed \pab/\ha\ line ratio.  
In exploring this possibility further, we found that a range of additional slit loss corrections 
($f_\mathrm{slit}$) are allowed by the data. 
In particular, for three out of the eleven galaxies the allowed range straddled 
$f_\mathrm{slit}=0$ and standard dust law slopes, suggesting that the applied slit loss corrections were likely correct. 
For six of the galaxies, however, additional slit loss corrections of at least 6-93\%, depending on the galaxy, 
were required in order for the line ratios to be consistent with acceptable dust law parameters. 
 
Since the nominal slit loss correction described in Section~\ref{sec:sample} assumed a 
constant emission line equivalent width across the galaxy (both inside and outside the slit), 
deviations from this assumption would lead to errors in the resulting flux ratios.  
This would be the case if the line emission is more or less extended than the 
continuum emission from the galaxy.  Indications from the 3D-HST survey are that 
this is the case at $z\sim1$, with \ha\ emission being somewhat more extended than 
the $R$-band continuum \citep{nelson2012,nelson2013}. 
In addition, recent results from the CLEAR survey suggest that 
the relative compactness of the line and continuum emission may correlate 
with galaxy mass (J. Matharu et al. 2021, in prep.).  
Specifically, for galaxies less massive than $\mathrm{log}(M_{*}/M_{\odot}) \sim 9.2$ (8/11 galaxies in the current sample), 
the \ha\ emission appears to be comparable to or more compact on average than the continuum, which 
would lead to a slit loss overestimate (negative $f_\mathrm{slit}$). 
The applicability of this result to our galaxy sample is unclear, for while this subset includes the three galaxies with negative 
allowed $f_\mathrm{slit}$, it also includes two galaxies for which only large positive $f_\mathrm{slit}$ values are allowed. 
On the other hand, three galaxies in the sample are more massive than $\mathrm{log}(M_{*}/M_{\odot}) \sim 9.2$, where there is evidence 
that \ha\ emission will on average be more extended than the continuum.  In this case, one would expect the slit loss correction 
to be underestimated (positive $f_\mathrm{slit}$) leading to enhanced \pab/\ha\ ratios. 
Indeed, all three of these more massive galaxies require large positive $f_{slit}$ values, 
and intriguingly, two of them (GN3 34456, GN3 34157) represent the highest two \pab/\ha\ ratios in the sample.  
Thus, while it is difficult to draw strong conclusions, it appears that variations in equivalent width across the galaxy, 
and the resulting underestimated slit loss corrections, could explain some of the most enhanced \pab/\ha\ line ratios.

Clearly, it would be best to circumvent these issues by removing the need for slit loss corrections altogether. 
This could be done either by using disk-integrated Balmer line fluxes from narrowband imaging 
(requiring a correction for \nii\ in the case of \ha), integral field spectroscopy (IFS), or 
slitless grism observations, or by obtaining all three line measurements with a consistent spectroscopic slit 
from space or with appropriate correction for atmospheric dispersion. 
In addition, high signal-to-noise constraints on all three lines will be important 
for deriving meaningful constraints on the dust law normalization and slope. 
From the existing data on this sample, one galaxy (GN2 19221) is in that position, where the effects 
of sub-unity covering fraction are likely negligible, the slit losses are plausibly 
well-accounted for, and the observational uncertainties are small enough that galaxy 
is constrained to have $n\sim1.3$ and $\tau_{V}\sim1$. This galaxy therefore provides a promising example of what will be possible in the near future.

\subsection{Galaxy Morphologies}
\begin{figure}
\center
\includegraphics[angle=0,width=5.5in]{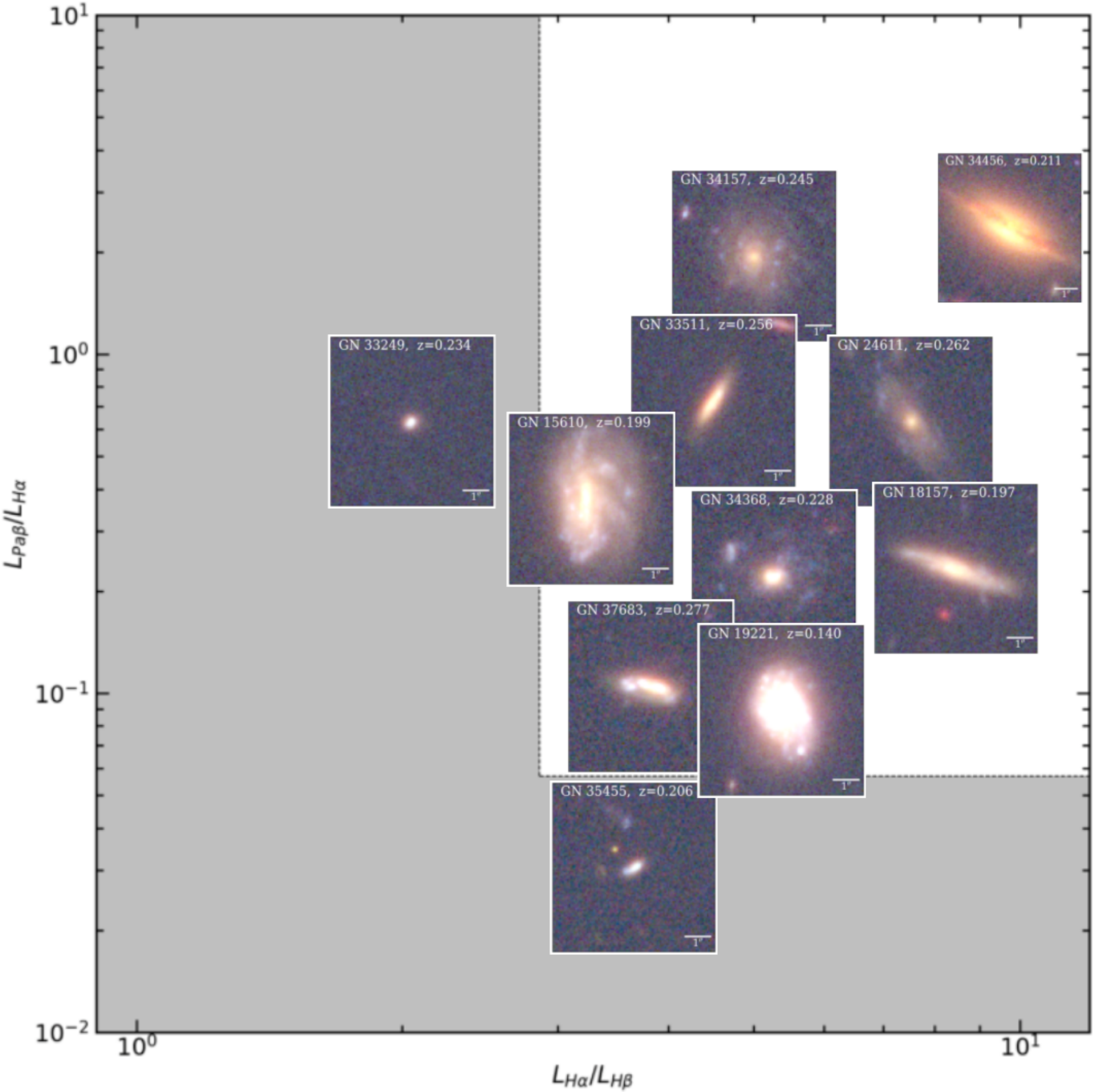}
\caption{Observed \pab/\ha\ and \ha/\hb\ ratios for the eleven galaxies in the \citet{cleri2020} 
sample, as shown in Figure~\ref{fig:fiduciallineratios} (right) but here indicated with $iYH$-band composite images. The intrinsic Case~B recombination line ratios are shown as dotted black lines. The sample includes a diverse range of galaxy morphologies but no obvious trends with line ratios. The highest \pab/\ha\ ratio in the sample, GN 34456, shows a significant dust lane through its center. 
}
\label{fig:galimages}
\end{figure}

It is natural to wonder whether the morphologies across the sample 
can explain some of the variation in recovered dust law parameters. 
To explore this, Figure~\ref{fig:galimages} replicates the \pab/\ha\ vs. \ha/\hb\ line ratio plot, 
but with each galaxy indicated using its composite CANDELS $iYH$-band imaging. 
However, no clear trends emerge.  
The four galaxies most consistent with standard dust law slopes (GN2 19221, GN1 37683, GN2 18157, GN3 34368)
show a range of morphologies with examples of knotty star-forming disks (GB2 19221, GB3 34368), 
one edge-on galaxy (GB2 18157), and one that resembles a ``chain'' galaxy (GN1 37683).
The effective radii for this subset vary widely ($r_{eff}$=0.3-1.20\arcsec; see Table 2). 
For the six galaxies at or near prior boundaries and the one with no solution, the morphologies also span a 
broad range, including knotty star-forming disks (GN2 15610, GN4 24611, GN3 34157), 
two that are very compact (GN5 33249, GN3 35455), and two edge-on systems (GN3 33511, GN3 34456), with the latter 
showing a particularly prominent dust lane.  
The effective radii span a similarly broad range ($r_{eff}=$0.2-1.49\arcsec). 
Thus, galaxy morphology is not obviously correlated with the dust law constraints within this sample.

\subsection{Non-universal Dust Laws}

In the existing analysis, we found that some galaxies are in agreement with standard dust laws, within 
the broad confidence intervals. In other cases, the solutions are skewed to steeper or shallower dust law 
slopes, and when treating the entire sample as a single composite galaxy, we found a preference for 
shallower dust law slopes.
Thus, taking the results from our relatively small sample of galaxies at face value, 
we see tentative evidence for a 
variety of dust laws in galaxies at $z\sim0.2$.  This would not be entirely surprising. 
Even in the local universe we see ample indications of variations, with the Milky Way 
dust attenuation curve differing substantially in shape from those of the neighboring Large and Small Magellanic Clouds 
\citep[LMC, SMC; e.g.,][]{pei1992,gordon2003}.  At higher redshift ($z\sim0.5-3$), 
SED-fitting approaches have presented evidence for the non-universality of 
the dust law in galaxies \citep{kriek2013,salmon2016}.  
This is perhaps to be expected, as the dust extinction law will reflect the dust 
composition and the grain size distribution, which will in turn reflect the dust 
production processes in the galaxy.  
Moreover, the geometry of the ISM can lead to different emergent dust attenuation curves 
in the optical and near-infrared even if the intrinsic extinction curves are the same, implying that 
attenuation curves aren't uniquely determined even if the stellar mass, star formation rate, 
and metallicity are known \citep{narayanan2018}.  

We have shown that constraining dust laws using more than 
two emission lines can, in principle, provide a nearly model-independent, complementary way to study the 
intrinsic variability in dust attenuation law parameters. 
In doing so, we have used the assumption of a power-law dust law. 
This was a simplification but was useful for exploring the three-line method 
along with various astrophysical and observational challenges. 
However, in reality the dust law will have a more complicated 
shape and potentially additional features, e.g., the 2175\AA\ bump. 
It is possible that some of the other galaxies in the sample could be better explained by a 
different dust law. 
It would be beneficial therefore to further investigate the detailed 
shape of the nebular dust law once more emission line constraints are 
available for a larger sample of galaxies. 

\subsection{Leveraging the Three-Line Approach}

This work motivates a number of additional investigations to leverage the potential power 
of multiple emission lines to put constraints on the shape of the dust law in galaxies. 
In the local universe, \pab\ measurements from HST already exist for a few galaxies 
\citep[e.g., NGC5194 and NGC6946;][]{kessler2020}.  
Narrowband imaging or, better, IFS observations of 
these galaxies could be used to study the dust law in individual apertures across the 
galaxy disks, while disk-integrated measurements could be compared with those derived 
at higher redshifts. 

Even more exciting is the prospect of using the James Webb Space Telescope (JWST) 
to follow up order-of-magnitude larger samples of intermediate to high-redshift galaxies 
(z$\approx$0.3-2), where at least three nebular lines can be observed 
using identical NIRSpec Microshutter Array (MSA) slitlets, the NIRSpec IFU, or the NIRISS and NIRCam grisms, 
removing the issue of differential slit losses between key lines. 
Using deeper near-infrared constraints on Paschen lines from JWST and a proper treatment of upper limits, 
we can obtain more representative and largely model-independent constraints on the slope and normalization 
of the dust law of galaxies at the peak of cosmic star formation. 

%%%%%%%%%%%%%%%%%%%%%%%%%%%%%%%%%

\section{Conclusions}
\label{sec:conclusions}

In this paper, we explored both the power of and practical challenges associated 
with using three hydrogen emission lines to constrain not just the normalization but also the 
slope of the dust attenuation law.  
Using a sample of eleven galaxies with existing \ha, \hb, and \pab\ 
measurements, and accounting for observational uncertainties, 
we showed that the dust law slope and normalization of one galaxy are well-constrained, an indication 
of what will be possible in future work. 
Eight of the other ten galaxies can be explained using a simple power-law dust attenuation 
curve given current observational uncertainties, although some prefer dust law slopes that are significantly 
steeper or shallower than expected from theoretical arguments. 
This may suggest that low-mass, low-redshift galaxies may have a diversity 
of dust laws, reminicent of the claims of varying dust laws in the higher redshift Universe.
However, we then explored other astrophysical or observational biases that may influence the 
derived results. 
Using an analytic approach, we showed that high but sub-unity ($>$97\%) covering fractions 
of dusty ISM can explain some of the unusual ratios seen in this sample of galaxies, but that different slit loss 
corrections between key lines may also be contributing to the results.
This work therefore emphasizes how the three-line approach can provide 
important complementary constraints on the shape of the dust attenuation law, 
but that a naive implementation can overlook potentially serious systematic uncertainties. 
With deeper measurements and a consistent observational set-up across all emission lines, 
we can hope to 
constrain the slope and normalization of the dust attenuation law, and potentially, with 
additional recombination line constraints, the dust covering fraction in individual galaxies.
This in turn will improve measurements of key galaxy properties including gas-phase metallicities, ionization 
parameters, and instantaneous star formation rates.

\section{acknowledgments}
The authors would like to thank Ben Weiner, Jasleen Matharu, and Syndey Lower for helpful discussions,
and the anonymous referee for useful suggestions that substantially improved the quality of this paper.
Figures 1-6 were created using color-blind-friendly IDL color tables developed by
Paul Tol (https://personal.sron.nl/~pault/). 
MKMP acknowledges support from NSF grant AAG-1813016, and KMF acknowledges support from NSF grant AAG-2006550.  
NJC and JRT acknowledge support from NSF grant CAREER-1945546 and NASA grant JWST-ERS-01345.

This work is based on data obtained for the CLEAR program (GO-14227),
the 3D-HST Treasury Program (GO 12177 and 12328), and
the CANDELS Multi-Cycle Treasury Program
by the NASA/ESA Hubble Space Telescope, which is operated by the
Association of Universities for Research in Astronomy, Incorporated, under NASA contract NAS5- 26555.

%Input any tables
\begin{deluxetable}{cccc}
\tabletypesize{\scriptsize}
%\rotate
\label{tab:coveringfraction}
\tablecaption{Dust Covering Fractions}
\tablewidth{0pt}
\tablehead{
\colhead{Field} & \colhead{ID} & \colhead{$f_\mathrm{cov,min}$} & \colhead{$f_\mathrm{cov,max}$} \\
}
\startdata
GN1 &        37683 &   0.64 &   0.99 \\ 
GN2 &        19221 &   0.64 &   1.00 \\ 
GN2 &        15610 &   0.84 &   0.97 \\ 
GN2 &        18157 &   0.89 &   1.00 \\ 
GN3 &        34456 &   0.99 &   1.00 \\ 
GN3 &        34157 &   0.97 &   0.98 \\ 
GN3 &        33511 &   0.93 &   0.99 \\ 
GN3 &        34368 &   0.72 &   1.00 \\ 
GN3 &        35455 & - & - \\ 
GN4 &        24611 &   0.96 &   1.00 \\ 
GN5 &        33249 & - & - \\ 
\enddata
\tablecomments{These are the minimum and maximum allowed dust covering fractions that lead to non-negative dust law slopes $n$ and normalizations $\tau_{V}$, and a star formation rate of SFR$<$10 M$_{\odot}$ yr$^{-1}$.}
\end{deluxetable}

\begin{deluxetable}{ccccc}
\tabletypesize{\scriptsize}
%\rotate
\label{tab:empiricalslitlosses}
\tablecaption{Empirical Slit Losses}
\tablewidth{0pt}
\tablehead{
\colhead{Field} & \colhead{ID} & $r_{eff}$\tablenotemark{a} & \colhead{Slit Loss} & \colhead{Slit Loss} \\
 & & (arcsec) & (best case) & (worst case) \\
}
\startdata
GN1 &        37683 &   0.69 &   0.01 &   0.36 \\ 
GN2 &        19221 &   0.60 &   0.27 &   0.37 \\ 
GN2 &        15610 &   1.49 &   0.45 &   0.68 \\ 
GN2 &        18157 &   1.20 &   0.09 &   0.56 \\ 
GN3 &        34456 &   1.10 &   0.32 &   0.56 \\ 
GN3 &        34157 &   1.10 &   0.49 &   0.60 \\ 
GN3 &        33511 &   0.60 &   0.00 &   0.27 \\ 
GN3 &        34368 &   0.30 &   0.13 &   0.22 \\ 
GN3 &        35455 &   0.22 &   0.00 &   0.03 \\ 
GN4 &        24611 &   1.30 &   0.37 &   0.60 \\ 
GN5 &        33249 &   0.20 &   0.01 &   0.02 \\ 
\enddata
\tablenotetext{a}{Effective radii ($r_{eff}$) are taken from \citet{vanderwel2012}.}
\tablecomments{Empirical best (worst) case slit losses represent the most (least) likely slit loss corrections applied to the tabulated Balmer fluxes. These were estimated using the observed stellar continuum morphology of each galaxy, the slit orientation, and the assumption that nebular and stellar continuum morphologies are equal, as described in the text.}
\end{deluxetable}

\begin{deluxetable}{ccccc}
\tabletypesize{\scriptsize}
%\rotate
\label{tab:slitlossunderestimates}
\tablecaption{Additional Allowed Slit Loss Correction}
\tablewidth{0pt}
\tablehead{
\colhead{Field} & \colhead{ID} & \colhead{$\mathrm{log}$ M$_{*}$\tablenotemark{a}} & \colhead{$f_\mathrm{slit,min}$} & \colhead{$f_\mathrm{slit,max}$} \\
 & & (M$_{\odot}$) & & \\
}
\startdata
GN1 &        37683 &       8.61 &   0.06 &   0.49 \\ 
GN2 &        19221 &       9.05 &  -0.40 &   0.40 \\ 
GN2 &        15610 &       9.43 &   0.77 &   0.81 \\ 
GN2 &        18157 &       8.96 &  -0.30 &   0.71 \\ 
GN3 &        34456 &      10.01 &   0.87 &   0.95 \\ 
GN3 &        34157 &       9.23 &   0.93 &   0.95 \\ 
GN3 &        33511 &       8.63 &   0.81 &   0.89 \\ 
GN3 &        34368 &       8.47 &  -0.12 &   0.53 \\ 
GN3 &        35455 &       7.69 &  -0.99 &  -0.71 \\ 
GN4 &        24611 &       8.91 &   0.51 &   0.87 \\ 
GN5 &        33249 &       7.67 & - & - \\ 
\enddata
\tablenotetext{a}{Stellar masses are taken from \citet{skelton14}.}
\tablecomments{These are the minimum and maximum allowed $f_\mathrm{slit}$ values (implying a previous slit loss correction overestimate if negative or underestimate if positive) that lead to non-negative dust law slopes $n$ and normalizations $\tau_{V}$, and a star formation rate of SFR$<$10 M$_{\odot}$ yr$^{-1}$.}
\end{deluxetable}

\clearpage

%Insert figures and captions
%\maxdeadcycles=1000

\end{document}